\documentclass[
amsmath,amssymb,
aps,
superscriptaddress
frontmatterverbose, 
]
{revtex4-2}
\usepackage{graphicx}% Include figure files
\usepackage{dcolumn}% Align table columns on decimal point
\usepackage{bm}% bold math
\usepackage{hyperref}% add hypertext capabilities
\usepackage{amsmath}
\usepackage{comment}
\usepackage{tikz}
\usepackage{subfigure}
\usepackage[nocomma]{optidef}
\usepackage{algorithm}
\usepackage{algpseudocode}
\begin{document}

\preprint{APS/123-QED}

\title{Identifying spatially-localized instability mechanisms using sparse optimization}

\author{Talha Mushtaq}
 \altaffiliation[Current address: ]{Gulfstream Aerospace Corporation, Savannah, GA 31408, USA}
\author{Maziar S. Hemati}

% \email{mhemati@umn.edu}
 %Lines break automatically or can be forced with \\
%
% \email[Corresponding author: ]{mhemati@umn.edu}
\affiliation{%
Aerospace Engineering \& Mechanics\\
 University of Minnesota\\ Minneapolis, MN 55455, USA
}%

\date{\today}% It is always \today, today,
             %  but any date may be explicitly specified

\begin{abstract}
Recent investigations have established the physical relevance of spatially-localized instability mechanisms in fluid dynamics and their potential for technological innovations in flow control.
In this letter, we show that the mathematical problem of identifying spatially-localized optimal perturbations that  maximize perturbation-energy amplification can be cast as a sparse (cardinality-constrained) optimization
problem.
Unfortunately, cardinality-constrained optimization problems are non-convex and combinatorially hard to solve in general.
To make the analysis viable within the context of fluid dynamics problems, we propose an efficient iterative method for computing sub-optimal spatially-localized perturbations.
Our approach is based on a generalized Rayleigh quotient iteration algorithm followed by a variational renormalization procedure that reduces the optimality gap in the resulting solution.
The approach is demonstrated on a sub-critical plane Poiseuille flow at $Re=4000$, which has been a benchmark problem studied in prior investigations on identifying spatially-localized flow structures.
Remarkably, we find that a subset of the
perturbations identified by our method yield a comparable degree of energy amplification as their global counterparts.
We anticipate our proposed analysis tools will facilitate further investigations into spatially-localized flow instabilities, including within the resolvent and input-output analysis frameworks. 
\end{abstract}

%Use showkeys class option if keyword
                              %display desired
\maketitle

\section{\label{sec:intro}Introduction}

Non-modal stability analysis is conducted by determining the maximum amplification a flow perturbation can exhibit over a finite time horizon~\cite{schmidBook}.
The associated optimal flow perturbations that result in the maximum transient growth (i.e.,~energy amplification) are global in nature, tending to involve many flow quantities distributed across the full spatial domain.
However, it is often of physical interest to consider optimal perturbations that are sparse and spatially-localized: identification of spatially-localized optimal perturbations would reveal specific regions of the spatial domain and the associated flow quantities that are most pertinent to triggering instabilities.
Recent investigations have shown that such information can guide the placement of actuators and inform the development of effective flow control strategies~\cite{skeneJFM2022,tamilselvamAIAA2022,bhattacharjeeTCFD2020}.

Spatially-localized stability analysis is naturally posed as a cardinality-constrained optimization problem, as will be shown in this letter.
However, cardinality-constrained optimization problems are notoriously difficult to solve due to the inherent need for a combinatorial search.
As such, prevailing methods relax the cardinality-constraint and instead work with a modified but tractable optimization problem that promotes sparsity in the resulting solution through some heuristic~\cite{foures2013localization,skeneJFM2022,lopezAIAA2023,heide2023optimization,heideARXIV2024}.
For example, sparsity-promoting $\ell_1$-norms have been used to sparsify optimal forcing modes in resolvent analysis~\cite{skeneJFM2022,lopezAIAA2023} and to sparsify optimal perturbations in nonlinear non-modal stability analysis~\cite{heide2023optimization,heideARXIV2024}.
Despite their successes, the drawback of employing sparsity-promoting heuristics is that such analyses do not guarantee a given degree of sparsity \emph{a priori}, and so will often require computationally expensive parameter sweeps to realize sparse and physically meaningful solutions.

In this letter, we revisit cardinality-constrained optimization as the natural setting for conducting 
spatially-localized instability analysis.
For brevity, this letter focuses on linear non-modal stability analysis (see Section~\ref{sec:formulation}), but the ideas developed are equally valid within the context of any analysis method based on generalized Rayleigh quotient maximization (e.g.,~resolvent and input-output analysis).
As noted earlier, cardinality-constrained optimization problems are non-convex and combinatorially hard to solve in general.
To make the analysis tractable, we propose an efficient iterative method for computing sub-optimal spatially-localized perturbations (see Section~\ref{sec:solution}).
Our approach employs two key steps: (i)~find a ``good'' sparsity pattern for a given cardinality $k$, followed by 
    (ii)~determine an optimal solution for that sparsity pattern.
    In particular, we formulate a generalized Rayleigh quotient iteration algorithm to achieve~(i), then apply a variational renormalization procedure to reduce the optimality gap in the solution to achieve~(ii).
The approach is demonstrated on a sub-critical plane Poiseuille flow at $Re=4000$, which has been a benchmark problem studied in prior investigations on identifying spatially-localized flow structures (see Section~\ref{sec:results}).
Remarkably, we find that a subset of the
perturbations identified by our method yield a comparable degree of energy amplification as their global counterparts.
Further, our proposed approach decisively reveals the well-known lift-up mechanism as a driver of transient growth: sparse and spatially localized optimal perturbations are comprised solely of the wall-normal component of velocity, which results in transient amplification of streamwise streaks.
Concluding remarks are given in Section~\ref{sec:conclusion}

\section{\label{sec:problem}Problem Formulation and Solution Approach}
\subsection{Problem Formulation}\label{sec:formulation}

Consider the linear dynamics of perturbations about a steady baseflow in the spatially discretized form,
\begin{equation}
    \frac{\mathrm{d}}{\mathrm{d}t}q(t) = L q(t)
\end{equation}
where $q\in\mathbb{C}^n$ denotes the state vector\footnote{Here we work with complex variables because we will work with operators formulated using spectral methods in later sections.
For real variables, the methods will be analogous to the ones derived here.}.
The state response due to an initial perturbation $q(t_0)=q_0$ at any instant $t\ge t_0$ can be determined as $q(t)=\Phi(t,t_0)q_0$, where $\Phi(t,t_0)$ is the state transition matrix.
Thus, given the knowledge of the state at any time $t$, the energy can be computed as
$E(t)=q(t)^HQq(t)$, where $(\cdot)^H$ denotes Hermitian transpose and $Q=Q^H>0$.
For example, when $E$ is the kinetic energy, then $Q$ would be an appropriate matrix of energy and quadrature weights.
The amplification of energy over a prescribed time interval can then be expressed in terms of the generalized Rayleigh quotient
\begin{equation}
R(q_0,P,Q)=\frac{q_0^HPq_0}{q_0^HQq_0}, 
\end{equation}
where $P=P(t,t_0):=\Phi^H(t,t_0)Q\Phi(t,t_0)$.
If $L^HQ+QL\nleq0$, then there will exist an initial condition $q_0$ that triggers a non-unity transient growth---i.e.,~a non-trivial amplification of energy $R(q_0,P,Q)=E(t)/E(t_0)>1$ for some $t>t_0$.
In non-modal stability analysis, the maximum amplification is of interest.
The \emph{optimal perturbation} $q_0^*$ that maximizes $R(q_0,P,Q)$ can be found by solving
the generalized eigenvalue problem
\begin{equation}
   Pq_0=\lambda Qq_0.
   \label{eq:GEVP}
\end{equation}
The corresponding maximum energy amplification $G(t)$ is simply
\begin{equation}
    G(t)=R(q_0^*,P,Q)=\lambda_\text{max} (P,Q),
    \label{eq:GRQmax}
\end{equation}
where $\lambda_\text{max}(P,Q)$ denotes the maximum generalized eigenvalue of the pair~$(P,Q)$.
We note that this generalized Rayleigh quotient maximization problem is equivalent to the following  quadratically-constrained quadratic program~(QCQP),
\begin{maxi}|l|[2]
{q_0}{q_0^HPq_0}{}{}
\addConstraint{q_0^HQq_0 = 1}.
\label{eq:nsp_gr}
\end{maxi}%

Further, the generalized eigenvalue problem results because we are aiming to maximize the generalized Rayleigh quotient $R(q_0,P,Q)$.
Using a simple change of variables $y(t)=Q^{1/2}q(t)$ and defining $S:=Q^{-1/2}PQ^{-1/2}>0$, we could instead find $G(t)$ by solving a standard eigenvalue problem.
In particular, $G(t)=R(y_0^*,S)$, where $y^*_0$ is the optimal perturbation represented in the transformed coordinates and $R(y_0,S):=R(y_0,S,I)$ is the standard Rayleigh quotient with $I$ denoting the $n\times n$ identity matrix.

In general, optimal perturbations $q_0^*$ are global in nature, i.e., the unit-energy associated with the perturbation is distributed among all state variables and therefore across all of the spatial domain.
In this work, we seek ``sparse'' solutions $q^{*}_{0_{sp}}$ that correspond to a subset of the state variables and tend toward spatially-localized perturbations---i.e.,~the perturbation energy is concentrated locally, rather than distributed globally.
To this end, we propose solving the cardinality-constrained QCQP,
\begin{maxi}|l|[2]
{q_0 \in \mathbb{C}^n}{q_0^HPq_0}{}{}
\addConstraint{q_0^HQq_0 = 1}
\addConstraint{\mathrm{card}(q_0)\le k},
\label{eq:optQCQP}
\end{maxi}%
which yields solutions that are at most $k$-sparse (i.e., the number of non-zero entries in $q_{0_{sp}}^{*}$ will be at most $k$).
For the case where $k=n$, the cardinality constraint becomes redundant and the optimization in \eqref{eq:optQCQP} simplifies to \eqref{eq:nsp_gr}.
However, when $k<n$, the optimization in \eqref{eq:optQCQP} is non-convex and the optimal solution requires a combinatorial search over all possible sparsity patterns.

The combinatorial nature of the solution to \eqref{eq:optQCQP} %and \eqref{eq:optS} 
makes optimal solutions intractable for most fluid dynamics applications due to the high-dimension of the state space.
However, for relatively small problem instances (e.g.,~$n\lesssim 30$) optimal solutions can be computed using combinatorial-optimization methods, e.g., dedicated branch-and-bound algorithms~\cite{moghaddam}. 
As such, various heuristics and convex relaxations are often used to obtain sub-optimal solutions.
For instance, it is common to incorporate sparsity-promoting $p$-norms---with $0<p\le1$---to regularize the optimization problem and obtain sparse solutions to a different---but now tractable---optimization problem.
However, methods based on sparsity-promoting norms generally do not guarantee a particular degree of sparsity \emph{a priori};
rather, the sparse and ``physically meaningful'' solutions are identified only after sweeping over a large grid of regularization parameters and solving each associated optimization problem.

To circumvent the issues inherent to the relaxed solution methods, we present a computationally efficient numerical algorithm that is simple to implement and converges to sub-optimal solutions of the cardinality-constrained QCQP in \eqref{eq:optQCQP}.
Any solution to \eqref{eq:optQCQP} is guaranteed \emph{a priori} to be $k$-sparse (i.e.,~$\mathrm{card}(q_0)=k$).
As we will see, the resulting sparse solutions tend toward spatially-localized perturbations that reveal pertinent flow physics and instability mechanisms.

Before we discuss our proposed solution approach, it is instructive to consider again the change of variables $y(t)=Q^{1/2}q(t)$ discussed above.
In this case, the equivalent  cardinality-constrained QCQP will be
\begin{maxi}|l|[2]
{y_0}{y_0^HSy_0}{}{}
\addConstraint{y_0^Hy_0 = 1}
\addConstraint{\mathrm{card}(Q^{-1/2} y_0)\le k}.
\label{eq:optS}
\end{maxi}
The representation of the QCQP in \eqref{eq:optS} explicitly highlights the fact that the cardinality constraint is imposed on a different set of variables than the set used in defining the energy norms, i.e.,~cardinality is imposed on ${q_0=Q^{-1/2}y_0}$, and not on $y_0$.
This is an important point to note because it explicitly shows that the underlying problem structure is \emph{not} equivalent to a sparse PCA (principle components analysis) problem for which $\mathrm{card}(y_0)\le k$, but  rather a sparse LDA (linear discriminant analysis) problem for which $\mathrm{card}(Q^{-1/2}y_0)\le k$~\cite{moghaddam}.
In the next section, we will describe how we can extend existing algorithms for sparse PCA to accommodate the special sparse LDA problem structure inherent to~\eqref{eq:optQCQP}.

\subsection{Solution Approach} \label{sec:solution}

We begin with the following observation:
For any $k$-sparse vector $q_{0_{sp}} \in \mathbb{C}^n$ and a given matrix pair $(P,Q)$,
\begin{equation}
%    \lambda = \frac{q_{0_{sp}}^{H} P q_{0_{sp}}}{q_{0_{sp}}^{H} Q q_{0_{sp}}} = \frac{z^H P_k z}{z^H Q_k z},
\lambda = R(q_{0_{sp}},P,Q)=R(z,P_k,Q_k)
    \label{eq:optsol}
\end{equation}
where $z \in \mathbb{C}^k$ is the sub-vector containing only the $k$ non-zero indices of $q_{0_{sp}}$ and $P_k \in \mathbb{C}^{k \times k}$ and $Q_k \in \mathbb{C}^{k \times k}$ are, respectively, the  principal minors obtained from deleting the rows and columns of $P$ and $Q$ associated with zero indices of $q_{0_{sp}}$.
%
%Given $Q > 0$, we can always normalize $q_{0_{sp}}$ such that $q_{0_{sp}}^H Q q_{0_{sp}} = 1$ and same follows for $z$.
%
Thus, the $z$ that maximizes $\lambda$ in \eqref{eq:optsol} will be the principal generalized eigenvector of the pair $(P_k, Q_k)$. %, i.e., $z^* = v_{\max}(P_k,Q_k)$.
%
% This is equivalent to maximizing the objective function given in \eqref{eq:optQCQP} by setting $z^H Q_k z = 1$ without loss of generality.
%
It follows that an optimal sub-vector $z^*$ obtained from \emph{any} method solving \eqref{eq:optQCQP} must satisfy $z^* = v_{\max}(P_k,Q_k)$ and consequently $\lambda^*=\lambda_{\max}(P_k, Q_k)$~\cite{moghaddam}. % will be the largest generalized eigenvalue~\cite{moghaddam}.
Therefore, computing an optimal sparse solution from \eqref{eq:optQCQP} inherently requires a combinatorial search over all possible sparsity patterns for $k$-sparse vectors $q_{0_{sp}}$.

Here we propose a two-step approach to solving ~\eqref{eq:optQCQP} that avoids a combinatorial search: %(i)~find a ``good'' sparsity pattern, then (ii)~determine an optimal solution for that sparsity pattern.
\begin{enumerate}
    \item[] (i)~find a ``good'' sparsity pattern for a given cardinality $k$, then 
    \item[] (ii)~determine an optimal solution for that sparsity pattern. % via variational renormalization.
\end{enumerate}
In principle, step (i) involves using any systematic (or \emph{ad-hoc}) approach to find a sparsity pattern.
A simple approach can be devised by noting that any (non-sparse) candidate solution $q_0$ can be ``sparsified'' to a prescribed cardinality $k$ via Euclidean projection $\mathbb{P}_k$ onto the set $\{\|q_0\|_0=k\}\cap\{\|Q^{1/2}q_0\|_2=1\}$.
This approach is simple and efficient because it amounts to setting the $(n-k)$ smallest entries in $q_0$ to zero, then renormalizing the resulting vector to have unit norm.
A straight-forward way to do this is to use the ``simple thresholding'' technique, which involves zeroing out $(n-k)$ components of $q_0$ with the smallest absolute values.
This technique is popular for sparse principal components analysis~(PCA) problems and avoids any combinatorial searches (see \cite{moghaddam} for details).
Of course, there is no guarantee that the resulting solution will possess a ``good'' sparsity pattern for optimizing the underlying problem; however, this would be a simple way of sparsifying existing solutions for (non-sparse) optimal perturbations $q_0^*$ determined from the generalized eigenvalue problem in~\eqref{eq:nsp_gr}.
Here, we propose an efficient numerical method inspired by the generalized Rayleigh quotient iteration~(GRQI) algorithm originally presented in~\cite{kuleshov2013} to solve sparse PCA problems.
We will introduce our extension of the GRQI method to the more general setting of sparse LDA momentarily.

Step~(ii) is central to our approach as it ensures that a sub-optimal solution to the original problem posed in~\eqref{eq:optQCQP} is determined.
The key observation we exploit in conducting step~(ii) is as follows:
Fixing the sparsity pattern on the decision variable $q_0$ in~\eqref{eq:optQCQP} fixes the $k$ working indices for the problem and the original cardinality constraint becomes redundant.
In this case, we can use knowledge of the working indices to define $q_{0_{sp}}$ and its associated sub-vector $z_0$.
It then follows that the optimal sparse perturbation $q_{0_{sp}}^{*}$ for $k$ working indices can be determined by maximizing the generalized Rayleigh quotient $R(z,P_k,Q_k)$, which is simply the largest generalized eigenvalue $\lambda_{\max}(P_k, Q_k)$.
This is sometimes referred to as \emph{variational renormalization} in the machine learning literature~\cite{moghaddam}.
An important feature of variational renormalization is that it can be used to reduce the optimality gap in candidate solutions to~\eqref{eq:optQCQP} derived from \emph{any} method, including those based on sparsity-promoting norms and problems for which the cardinality constraint is imposed on a different variable.
In addition, maximizing $R(z,P_k,Q_k)$ in this step is computationally inexpensive because we work with the reduced matrices $P_k$ and $Q_k$.
We emphasize here that this variational renormalization approach works just as well in reducing the optimality gap in problems of sparse resolvent and input-output analysis~\cite{skene,lopez}.

It is important to note that the variational renormalization in step~(ii) yields an optimal solution for a given sparsity pattern identified in step~(i).
The \emph{generalized inclusion principle}~\cite{moghaddam} can be used to derive upper- and lower-bounds on the maximum amplification for $k$-sparse perturbations, regardless of the sparsity pattern.
The bounds are given in terms of the generalized eigenvalues as,
\begin{equation}
    \lambda_k(P,Q)\le \lambda_\text{max}(P_k,Q_k)\le \lambda_n(P,Q),
    \label{eq:GIP}
\end{equation}
where $\lambda_{k}$ is the $k$-th smallest generalized eigenvalue of $(P,Q)$ and $\lambda_n(P,Q)$ is the largest eigenvalue associated with the maximum amplification in the non-sparse setting.
The upper-bound establishes that $k$-sparse perturbations cannot induce a larger amplification than their non-sparse counterparts over a given time horizon $t=T$.
The lower-bound can guide the range of $k$ to grid over when pursuing sparse optimal perturbations that are able to impart a transient growth greater than a given threshold.

Returning to step~(i), we will now seek ``better'' solutions than might be obtained from the simple Euclidean projection $\mathbb{P}_k$ of the non-sparse $q_0$ followed by a variational renormalization.
To this end, we propose a further generalization of the generalized Rayleigh quotient iteration~(GRQI) of~\cite{kuleshov} in order to solve sparse LDA problems given by the QCQP in~\eqref{eq:optQCQP}.
We will refer to this generalization as a modified generalized Rayleigh quotient~(mGRQI) algorithm so as to distinguish it from the original GRQI algorithm from~\cite{kuleshov}.
As the name suggests, the original GRQI method generalizes the standard Rayleigh quotient iteration~(RQI) algorithm~\cite{parlett} to find sparse maximizers of the Rayleigh quotient $R(y_0,S)=R(y_0,S,I)$.
The GRQI method is attractive for fluids applications because it inherits the rapid (cubic) convergence of RQI and requires only modest computational effort when $k\ll n$, which tends to be the case when seeking sparse and spatially-localized perturbations:
The number of floating point operations per iteration scales as $\sim\mathcal{O}(k^3+nk)$~\cite{kuleshov}.
This is in contrast to generalized power methods~(GPM)~\cite{journee2010}---sparsity promoting variants of the standard power iteration---for which convergence is linear and the number of floating point operations per iteration scales as $\sim\mathcal{O}(n^2+nk)$.

The original GRQI algorithm solves the sparse PCA problem, but not the more general sparse LDA problem in~\eqref{eq:optQCQP}.
%
% %
Recall that the distinction is most apparent in the equivalent QCQP in~\eqref{eq:optS}, which is obtained through the change of variables $y_0 = Q^{1/2} q_0$:
in particular, the cardinality constraint is imposed on the inverse bijection of $y_0$, i.e., $\mathrm{card}(Q^{-1/2} y_0) = k$.
This constraint is difficult to satisfy short of a combinatorial search since it requires that $z_0 = Q_k^{-1/2} v_0$.
% %
However, we can use this relation between $z_0$ and $v_0$ to estimate a ``good'' working set for $k$ indices of $q_0$ within our modified GRQI~(mGRQI) algorithm, as described below. 
During the course of our investigation, we found that our proposed mGRQI algorithm yielded ``better'' sparsity patterns than either a simple Euclidean projection $\mathbb{P}_k$ of the non-sparse optimal $q_0^*$ or application of the standard GRQI---in the sense that variational renormalization tended to result in larger transient growth in an overwhelming majority of cases investigated.

% %
The implementation of mGRQI uses the matrix pair $(P,Q)$ to find a $k$-sparse $q_0$, denoted as $q_{0_{sp}}$.
Specifically, mGRQI performs a Rayleigh update on the set of $k$ non-zero indices of $q_0$ for a fixed sparsity pattern followed by a power update to obtain improvement in the working set for $q_{0}$ at each iterate.
The power update is similar in nature to the one used in the original GRQI; however, mGRQI performs the update on the full indices of $y_0$ instead of $q_0$ to maintain numerical stability and improve solution convergence.
Ideally, using $Q^{-1} P$ for the power update makes sense since it avoids transforming the system into $y_0$ coordinates and $q_0$ is directly related to the principal generalized eigenvector of $Q^{-1} P$.
However, $Q^{-1} P$ is not generally a self-adjoint operator, which can cause solutions to oscillate and introduce numerical errors.
The resulting $y_0$ from the power update is projected using $\mathbb{P}_k$ to get $y_{0_{sp}} = \mathbb{P}_k(y_0)$ and consequently, its associated sub-vector $v_0$ containing $k$ non-zero indices.
These indices can then be mapped back to the $k$ indices of $q_0$ using the relation $z_0 = Q_k^{-1/2} v_0$.
The resulting $z_0$ vector can be used to perform a variational renormalization to satisfy \eqref{eq:optsol} and obtain $q_{0_{sp}}$. 
The detailed mGRQI algorithm is reported as Algorithm~\ref{alg:GRQI}, and the combined two-step algorithm for mGRQI with variational renormalization is reported as Algorithm~\ref{alg:GRQI_VaR}.
%%%%%%%%%%%%%%%%%%%%%%%
%%%%%%%%%%%%%%%%%%%%%%%
\begin{algorithm}[H]
\centering
\caption{mGRQI($P,\,Q,\,q,\,J,\, k,\,\epsilon$)}\label{alg:GRQI}
\footnotesize
\begin{algorithmic}[1]
    \State Initialize $j \leftarrow 0$, $\lambda^{(j)} \leftarrow \lambda_{\max}(P,Q)$, $q^{(j)}_0 \leftarrow \mathbb{P}_k(q)$
    \Repeat 
    \State $\mathcal{W} \leftarrow \{ i| q^{(j)}_{0_i} \neq 0\}$
    %\State $y_{0_{\mathcal{W}}}^{(j)} = Q^{1/2}_{\mathcal{W}} q_{0_{\mathcal{W}}}^{(j)}$
    %\State $S_k = Q^{-1/2}_{\mathcal{W}} P_{\mathcal{W}} Q^{-1/2}_{\mathcal{W}}$
    \State $q_{0_{\mathcal{W}}}^{(j)} \leftarrow \left( P_{\mathcal{W}} - \lambda^{(j)} Q_{\mathcal{W}}\right)^{-1} q_{0_{\mathcal{W}}}^{(j)}$ (Rayleigh Update)
    \State $q_{0_{new}} \leftarrow \frac{q_{0}^{(j)}}{\|Q^{1/2} q_{0}^{(j)}\|_2} $
    \State $y_{0_{new}} \leftarrow Q^{1/2} q_{0_{new}}$
    \If{$j < J$}
         \State $y_{0_{new}} \leftarrow \left(Q^{-1/2} P Q^{-1/2}\right) y_{0_{new}}$ (Power Update)
    \EndIf
    \State $y^{(j)}_0 \leftarrow \mathbb{P}_k(y_{0_{new}})$
    \State $\mathcal{W}_y \leftarrow \{ i| y^{(j)}_{0_i} \neq 0\}$
    \State $q^{(j + 1)}_{0_{\mathcal{W}_y}} = Q_{\mathcal{W}_y}^{-1/2} y^{(j)}_{0_{\mathcal{W}_y}}$
    \State $q^{(j + 1)}_0 \leftarrow \frac{q^{(j + 1)}_0}{\|Q^{1/2} q^{(j + 1)}_0\|_2}$
    \State $\lambda^{(j+1)} \leftarrow \left(q_0^{(j+1)}\right)^{H} P q^{(j+1)}_0$
    \State $j \leftarrow j + 1$
    \Until{$\|q^{(j)}_0 - q^{(j - 1)}_0\|_2 < \epsilon$}
    \State \Return $q^{(j)}_0$
\end{algorithmic}
\end{algorithm}
%%%%%%%%%%%%%%%%%%%%%%%%
%%%%%%%%%%%%%%%%%%%%%%%%
\begin{algorithm}[H]
\centering
\caption{mGRQI-Variational Renormalization($P,\,Q,\,J,\,k,\,\epsilon$)}\label{alg:GRQI_VaR}
\footnotesize
\begin{algorithmic}[1]
    \State $q \leftarrow \mathrm{largest\, column\, of\,} Q^{-1} P$
    \State $q_0 \leftarrow \mathrm{mGRQI}(P,Q,q,J,k,\epsilon)$
    \State $\mathcal{W} \leftarrow \{ i| q_{0_i} \neq 0\}$
    \If{$q_{0_{\mathcal{W}}} \neq v_{\max}(P_{\mathcal{W}}, Q_{\mathcal{W}})$}
        \State $q_{0_{\mathcal{W}}} \leftarrow v_{\max}(P_{\mathcal{W}}, Q_{\mathcal{W}})$
    \EndIf
\end{algorithmic}
\end{algorithm}

\section{Results}\label{sec:results}
\subsection{Model Problem: Plane Poiseuille Flow}
We will use the incompressible plane Poiseuille flow model as an example to demonstrate our proposed analysis framework.
This is a canonical flow model that has been studied in prior works on stability analysis, including recent techniques investigating methods that promote spatial-localization of instability mechanisms~\cite{skeneJFM2022,lopez,foures2014optimal}.
Plane Poiseuille flow is a pressure-driven flow between two infinite parallel plates separated a distance $2h$ apart.
The steady base flow is a parabolic profile in the streamwise component of velocity $U(y)=1-y^2$, where $y\in[-1,1]$ is the non-dimensional coordinate in the wall-normal direction and $U(y)$ has been normalized by the centerline velocity.
The Reynolds number $Re=Uh/\nu$ is based on the channel half-height $h$, centerline velocity $U=U(y=0)$, and kinematic viscosity $\nu$.
We next consider the dynamics of perturbations $(u,v,w,p)$ about this baseflow.
No slip boundary conditions are imposed at both walls.
Since the flow is incompressible, we can equivalently express the dynamics in terms of wall-normal velocity perturbations $v(x,y,z,t)$ and wall-normal vorticity perturbations $\eta(x,y,z,t)$.
Homogeneous Dirichlet boundary conditions are imposed on $\eta$ at the walls, while homogeneous Dirichlet and Neumann boundary conditions are imposed on $v$ at the walls.  
Spatial Fourier expansions are applied in the streamwise ($x$) and spanwise ($z$) directions with associated wavenumbers $\alpha$ and $\beta$, respectively, such that $v(x,y,z,t)=\hat{v}_{\alpha,\beta}(y,t)e^{(i\alpha x + i\beta z)}$ and $\eta(x,y,z,t) =\hat{\eta}_{\alpha,\beta}(y,t)e^{(i\alpha x + i\beta z)}$.
Retaining only linear terms (i.e., considering the dynamics of infinitesimal perturbations about the baseflow) yields the familiar Orr-Sommerfeld and Squire equations.
Discretizing in the wall-normal direction using Chebyshev polynomials yields the spatially-discretized form,
\begin{equation}
    \underbrace{\frac{\mathrm{d}}{\mathrm{d}t}\begin{bmatrix}\hat{v}\\ \hat{\eta}\end{bmatrix}}_{\dot{q}} = \underbrace{\begin{bmatrix}L_{OS}&0\\ L_C&L_{SQ}\end{bmatrix}}_{L} \underbrace{\begin{bmatrix}\hat{v}\\ \hat{\eta}\end{bmatrix}}_{q}
    \label{eq:PPF}
\end{equation}
where 
$L_{OS}=M^{-1}\left(-i\alpha UM - i\alpha U'' - M^2/Re\right)$, $L_{SQ}=-i\alpha U - M/Re$, $L_C=-i\beta U'$, $M=(\alpha^2+\beta^2-D^2)$, and $D$ is the Chebyshev differentiation matrix.
The associated perturbation kinetic energy is computed as $E(t)=q^H(t)Qq(t)$, where $Q$ is the integration weight obtained from computing the Clenshaw-Curtis quadrature weights for the kinetic energy integral.

The flow state $q(t)=(\hat{v}(t), \hat{\eta}(t)) \in \mathbb{C}^{2N}$ is discretized in the wall-normal direction $y \in [-1,\, 1]$ using Chebyshev polynomials~\cite{schmidBook}, %\cite{trfethenSci1993},
where $N$ is the number of collocation points.
In this study, we select $N = 100$ as it provides sufficient accuracy in capturing the transient energy growth: specifically, we observe ${< 0.1 \%}$ change in (non-sparse) transient growth when $N$ is doubled beyond this value).
Since the objective function in \eqref{eq:optQCQP} is dependent on $t$, we solve the optimization problem by sweeping over $N_T$ values of time horizons $t=T$.
We use a linearly-spaced grid of $N_T = 101$ values for the time horizons in the interval $[10,\, 30]$, i.e., the values increment by $0.2$.

\subsection{Spatially-localized optimal perturbations and transient growth analysis}
\begin{figure}[hbt!]
\centering
\includegraphics[width=0.8\textwidth]{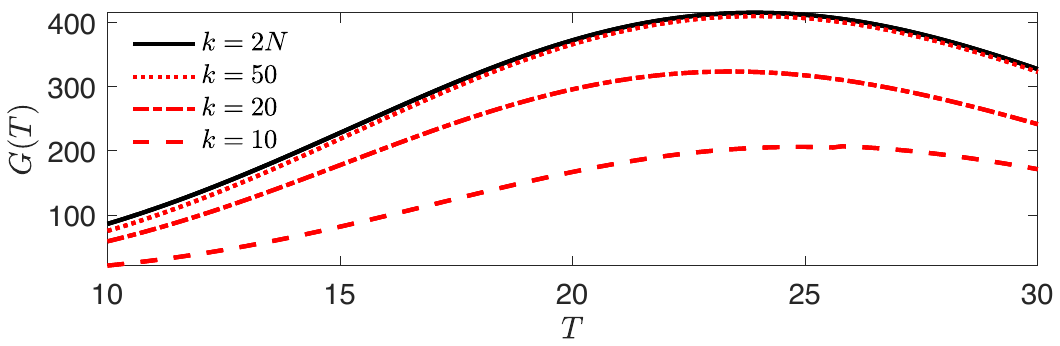}

\caption{Transient growth in kinetic energy $G(t=T)$ for a specified time-horizon $T$ for plane Poiseuille flow at $Re = 4000$ with $(\alpha,\beta)=(1,2)$ due to sparse optimal perturbations (red) approach that of the (non-sparse, $k=2N=200$) optimal perturbation as $k$ is increased.  At $k=50$, sparse optimal perturbations yield comparable transient growth for all time-horizons, including the maximum transient growth over all time-horizons $G=G_\text{max}$.  Associated optimal perturbations and responses are reported in figure~\ref{fig:composite_fields}.}
\label{fig:splinop}
\end{figure}

Sparse optimal perturbations $q_{0_{sp}}$ are computed using Algorithm \ref{alg:GRQI_VaR} for the linearized plane Poiseulle flow~\eqref{eq:PPF}.
In this letter, we only report results for the case $Re = 4000$, $\alpha = 1$, and $\beta = 2$, which are representative of our findings for other wave number pairs and Reynolds numbers.
The algorithm parameters are set as $J = \infty$ and $\epsilon = 10^{-6}$ for the power-method and the stopping criterion, respectively.
Sparse optimal perturbations were computed for $k=\{1,2,\dots,100\}$.
Here, we report and closely analyze three sets of results that highlight interesting properties of the algorithm and the flow physics---namely, $k = 10, 20,\, \mathrm{and}\, 50$.
The transient growth $G(t=T)$ is reported in figure~\ref{fig:splinop}.
The solid black curve is associated with $G(T)$ for (non-sparse) optimal perturbations for which $k=2N=200$, while each red curve corresponds to the $G(T)$ associated with a sparse optimal perturbation with $k=\{10,20,50\}$.
The relative magnitude of transient growth for different cardinalities $k$ follows from
the generalized inclusion principle~\eqref{eq:GIP} as expected, with the non-sparse $G(T)$ serving as an upper-bound to transient growth due to any sparse optimal perturbation.
As $k\rightarrow2N$, $G(T)$ for the sparse and non-sparse cases will be equivalent.
Strikingly, sparse optimal perturbations with $k=50<2N=200$ yield $G(T)$ within $<13\%$ of the non-sparse transient growth response for all $T$ considered.
This suggests that only a sub-set of flow quantities and/or spatial locations are essential to initiating the transient growth that is characteristic to this flow.
We will discuss the perturbations and the associated response in more detail momentarily.
When $k=50$, the maximum transient growth $G_{max}=G(T^*)=\max_T G(T)$---i.e.,~the peak in the transient growth curve---also occurs at the same time $T^*=24.0$ as in the non-sparse case.
The relative error between $G_{max}$ for $k=50$ and $k=2N=200$ is $<1.3\%$.
Interestingly, for $k=\{10,20\}$, the maximum transient growth $G_{max}$ occurs at a different time $T^*$ than in the non-sparse case.
For $k=20$, the maximum transient growth leads the non-sparse case ($T^*=23.4<24.0$); whereas, for $k=10$, the maximum transient growth lags the non-sparse case $T^*=25.8>24.0$.

\begin{figure}[hbt!]
 \centering
 \includegraphics[width=0.75\textwidth]{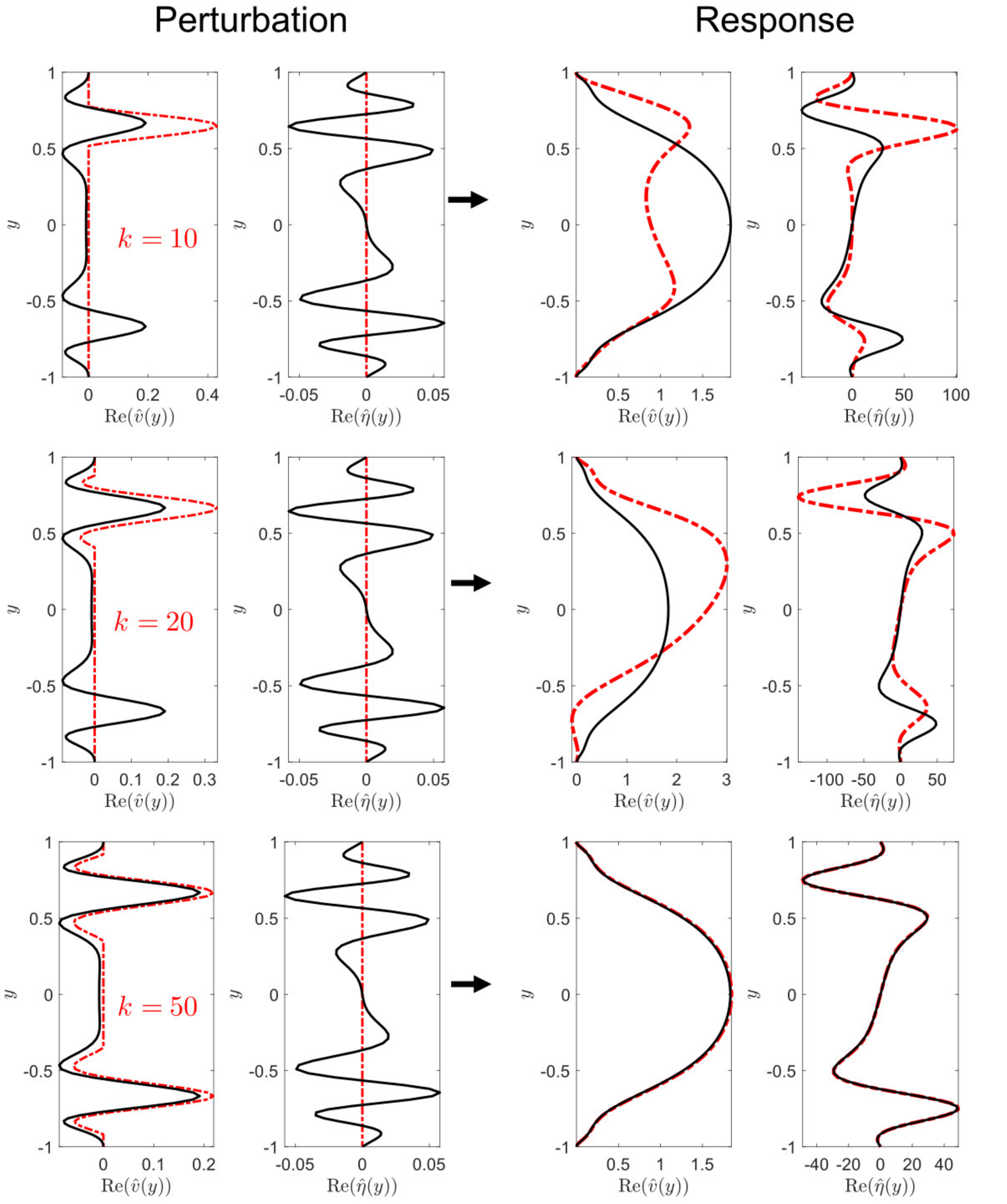}
 \caption{Optimal perturbations (left) and associated responses (right) resulting in $G=G_\text{max}$ for $Re=4000$ and $(\alpha,\beta)=(1,2)$. %Different components of $(\hat{v},\hat{\eta})$ are highlighted in the figure: Absolute values are solid lines (\full), real parts are dashed lines (\dashed) and imaginary parts are dotted lines (\dotted).
 Non-sparse optimal perturbations (black) are composed of wall-normal velocity $v$ and wall-normal vorticity $\eta$ distributed across the wall-normal direction ($y$). Sparse (sub-)optimal perturbations (red) with $k \le 50$ are composed only of wall-normal velocity.   %As $k$ is increased, the response converges to that of the non-sparse optimal perturbation.}
 }
 \label{fig:composite_fields}
\end{figure}

Figure~\ref{fig:composite_fields} reports optimal perturbations (left) and associated responses (right) corresponding with $G_{max}=G(T^*)$.
The black solid curves correspond to the non-sparse analysis ($k=2N=200$), whereas the red curves correspond to the cardinality-constrained transient growth analysis ($k=\{10,20,50\}$).
Notably, the non-sparse optimal perturbations include contributions from both wall-normal velocity~($\hat{v}$) and wall-normal vorticity~$\hat{\eta}$, and both quantities are ``active'' across the entirety of the wall-normal direction ($y$).
In contrast, all three of the sparse optimal perturbations reported only include contributions from wall-normal velocity, with absolutely no contribution from the wall-normal vorticity.
{\color{black}{These sparse solutions reveal a lift-up mechanism that commonly arises in sub-critical transition in shear flows: perturbations are comprised solely of the wall-normal component of velocity, which results in transient amplification of streamwise streaks in the response.}}
We note that since an optimal perturbation is defined to have unit energy, this fact results in the sparse optimal perturbations having an apparently larger magnitude of $\hat{v}$ than the non-sparse perturbation, which consequently distributes some of the unit energy into $\hat{\eta}$.
The sparsest solution reported ($k=10$) highlights the capability of the proposed approach for achieving spatially-localized optimal perturbations.
For the cases of $k=10$ and $k=20$, the localized $\hat{v}(y)$ structure reported favors proximity to the upper-wall; however, a symmetric and equally optimal sparse perturbation also exists near the lower-wall and was identified by our method by perturbing the initial iterate, but is not reported here.
As $k$ is increased from $k=10$ to $k=50$, the localized structure distributes further across the domain.
In the case of $k=50$, the sparse optimal perturbation profile for $\hat{v}$ closely resembles that of the non-sparse optimal perturbation.
Remarkably, the flow response at time $t=T^*$ shown in the figure is visually indistinguishable between $k=50$ and $k=2N=200$.
This striking similarity is consistent with the observation that $G_{max}$ for $k=50$ has a relative error of $<1.3\%$ with respect to the non-sparse case.

\section{Conclusions}\label{sec:conclusion}
In this letter, we proposed an iterative method for identifying sparse and spatially-localized instability mechanisms as sub-optimal solutions to a cardinality-constrained QCQP.
In contrast to prevailing sparsity-promoting methods, the proposed approach determines sub-optimal perturbations with a prescribed degree of sparsity, thereby removing the need for computationally demanding parameter sweeps.
In our study, the approach successfully identified spatially-localized perturbations in a plane Poiseuille flow at $Re=4000$ that achieved comparable transient growth as their global counterparts.
By increasing the cardinality of the solution, the method systematically revealed a lift-up instability mechanism.
While this letter focused on linear non-modal stability analysis, the proposed methods are equally applicable to related methods for stability analysis based on generalized Rayleigh quotient maximization (e.g.,~resolvent and input-output analysis).
We anticipate that our proposed algorithms will facilitate further research into sparse and spatially-localized instabilities in the future.

\begin{acknowledgments}
This material is based upon work supported by the Air Force Office of Scientific Research under award number FA9550-21-1-0434, the Office of Naval Research under award number N00014-22-1-2029, and the National Science Foundation under award number CBET-1943988.
\end{acknowledgments}

\bibliography{refs}% Produces the bibliography via BibTeX.

%apsrev4-2.bst 2019-01-14 (MD) hand-edited version of apsrev4-1.bst
%Control: key (0)
%Control: author (8) initials jnrlst
%Control: editor formatted (1) identically to author
%Control: production of article title (0) allowed
%Control: page (0) single
%Control: year (1) truncated
%Control: production of eprint (0) enabled
\providecommand{\noopsort}[1]{}\providecommand{\singleletter}[1]{#1}%
\begin{thebibliography}{17}%
\makeatletter
\providecommand \@ifxundefined [1]{%
 \@ifx{#1\undefined}
}%
\providecommand \@ifnum [1]{%
 \ifnum #1\expandafter \@firstoftwo
 \else \expandafter \@secondoftwo
 \fi
}%
\providecommand \@ifx [1]{%
 \ifx #1\expandafter \@firstoftwo
 \else \expandafter \@secondoftwo
 \fi
}%
\providecommand \natexlab [1]{#1}%
\providecommand \enquote  [1]{``#1''}%
\providecommand \bibnamefont  [1]{#1}%
\providecommand \bibfnamefont [1]{#1}%
\providecommand \citenamefont [1]{#1}%
\providecommand \href@noop [0]{\@secondoftwo}%
\providecommand \href [0]{\begingroup \@sanitize@url \@href}%
\providecommand \@href[1]{\@@startlink{#1}\@@href}%
\providecommand \@@href[1]{\endgroup#1\@@endlink}%
\providecommand \@sanitize@url [0]{\catcode `\\12\catcode `\$12\catcode
  `\&12\catcode `\#12\catcode `\^12\catcode `\_12\catcode `\%12\relax}%
\providecommand \@@startlink[1]{}%
\providecommand \@@endlink[0]{}%
\providecommand \url  [0]{\begingroup\@sanitize@url \@url }%
\providecommand \@url [1]{\endgroup\@href {#1}{\urlprefix }}%
\providecommand \urlprefix  [0]{URL }%
\providecommand \Eprint [0]{\href }%
\providecommand \doibase [0]{https://doi.org/}%
\providecommand \selectlanguage [0]{\@gobble}%
\providecommand \bibinfo  [0]{\@secondoftwo}%
\providecommand \bibfield  [0]{\@secondoftwo}%
\providecommand \translation [1]{[#1]}%
\providecommand \BibitemOpen [0]{}%
\providecommand \bibitemStop [0]{}%
\providecommand \bibitemNoStop [0]{.\EOS\space}%
\providecommand \EOS [0]{\spacefactor3000\relax}%
\providecommand \BibitemShut  [1]{\csname bibitem#1\endcsname}%
\let\auto@bib@innerbib\@empty
%</preamble>
\bibitem [{\citenamefont {Schmid}\ and\ \citenamefont
  {Henningson}(2001)}]{schmidBook}%
  \BibitemOpen
  \bibfield  {author} {\bibinfo {author} {\bibfnamefont {P.~J.}\ \bibnamefont
  {Schmid}}\ and\ \bibinfo {author} {\bibfnamefont {D.~S.}\ \bibnamefont
  {Henningson}},\ }\href@noop {} {\emph {\bibinfo {title} {Stability and
  Transition in Shear Flows}}}\ (\bibinfo  {publisher} {Springer–Verlag},\
  \bibinfo {address} {New York},\ \bibinfo {year} {2001})\BibitemShut {NoStop}%
\bibitem [{\citenamefont {Skene}\ \emph
  {et~al.}(2022{\natexlab{a}})\citenamefont {Skene}, \citenamefont {Yeh},
  \citenamefont {Schmid},\ and\ \citenamefont {Taira}}]{skeneJFM2022}%
  \BibitemOpen
  \bibfield  {author} {\bibinfo {author} {\bibfnamefont {C.}~\bibnamefont
  {Skene}}, \bibinfo {author} {\bibfnamefont {C.}~\bibnamefont {Yeh}}, \bibinfo
  {author} {\bibfnamefont {P.}~\bibnamefont {Schmid}},\ and\ \bibinfo {author}
  {\bibfnamefont {K.}~\bibnamefont {Taira}},\ }\bibfield  {title} {\bibinfo
  {title} {Sparsifying the resolvent forcing mode via gradient-based
  optimisation},\ }\href@noop {} {\bibfield  {journal} {\bibinfo  {journal}
  {Journal of Fluid Mechanics}\ }\textbf {\bibinfo {volume} {944}} (\bibinfo
  {year} {2022}{\natexlab{a}})}\BibitemShut {NoStop}%
\bibitem [{\citenamefont {Tamilselvam}\ \emph {et~al.}(2022)\citenamefont
  {Tamilselvam}, \citenamefont {Asztalos},\ and\ \citenamefont
  {Dawson}}]{tamilselvamAIAA2022}%
  \BibitemOpen
  \bibfield  {author} {\bibinfo {author} {\bibfnamefont {P.}~\bibnamefont
  {Tamilselvam}}, \bibinfo {author} {\bibfnamefont {K.}~\bibnamefont
  {Asztalos}},\ and\ \bibinfo {author} {\bibfnamefont {S.}~\bibnamefont
  {Dawson}},\ }\bibfield  {title} {\bibinfo {title} {Transient growth analysis
  of flow over an airfoil for identifying high-amplification,
  spatially-localized inputs},\ }\href@noop {} {\bibfield  {journal} {\bibinfo
  {journal} {AIAA Paper}\ } (\bibinfo {year} {2022})}\BibitemShut {NoStop}%
\bibitem [{\citenamefont {Bhattacharjee}\ \emph {et~al.}(2020)\citenamefont
  {Bhattacharjee}, \citenamefont {Klose}, \citenamefont {Jacobs},\ and\
  \citenamefont {Hemati}}]{bhattacharjeeTCFD2020}%
  \BibitemOpen
  \bibfield  {author} {\bibinfo {author} {\bibfnamefont {D.}~\bibnamefont
  {Bhattacharjee}}, \bibinfo {author} {\bibfnamefont {B.}~\bibnamefont
  {Klose}}, \bibinfo {author} {\bibfnamefont {G.}~\bibnamefont {Jacobs}},\ and\
  \bibinfo {author} {\bibfnamefont {M.}~\bibnamefont {Hemati}},\ }\bibfield
  {title} {\bibinfo {title} {Data-driven selection of actuators for optimal
  control of airfoil separation},\ }\href@noop {} {\bibfield  {journal}
  {\bibinfo  {journal} {Theoretical and Computational Fluid Dynamics}\ }\textbf
  {\bibinfo {volume} {34}} (\bibinfo {year} {2020})}\BibitemShut {NoStop}%
\bibitem [{\citenamefont {Foures}\ \emph {et~al.}(2013)\citenamefont {Foures},
  \citenamefont {Caulfield},\ and\ \citenamefont
  {Schmid}}]{foures2013localization}%
  \BibitemOpen
  \bibfield  {author} {\bibinfo {author} {\bibfnamefont {D.}~\bibnamefont
  {Foures}}, \bibinfo {author} {\bibfnamefont {C.}~\bibnamefont {Caulfield}},\
  and\ \bibinfo {author} {\bibfnamefont {P.~J.}\ \bibnamefont {Schmid}},\
  }\bibfield  {title} {\bibinfo {title} {Localization of flow structures using
  $\infty$-norm optimization},\ }\href@noop {} {\bibfield  {journal} {\bibinfo
  {journal} {Journal of Fluid Mechanics}\ }\textbf {\bibinfo {volume} {729}},\
  \bibinfo {pages} {672} (\bibinfo {year} {2013})}\BibitemShut {NoStop}%
\bibitem [{\citenamefont {Lopez-Doriga}\ \emph {et~al.}(2023)\citenamefont
  {Lopez-Doriga}, \citenamefont {Ballouz}, \citenamefont {Bae},\ and\
  \citenamefont {Dawson}}]{lopezAIAA2023}%
  \BibitemOpen
  \bibfield  {author} {\bibinfo {author} {\bibfnamefont {B.}~\bibnamefont
  {Lopez-Doriga}}, \bibinfo {author} {\bibfnamefont {E.}~\bibnamefont
  {Ballouz}}, \bibinfo {author} {\bibfnamefont {H.~J.}\ \bibnamefont {Bae}},\
  and\ \bibinfo {author} {\bibfnamefont {S.~T.}\ \bibnamefont {Dawson}},\
  }\bibfield  {title} {\bibinfo {title} {A sparsity-promoting resolvent
  analysis for the identification of spatiotemporally-localized amplification
  mechanisms},\ }\href@noop {} {\bibfield  {journal} {\bibinfo  {journal} {AIAA
  Paper}\ } (\bibinfo {year} {2023})}\BibitemShut {NoStop}%
\bibitem [{\citenamefont {Heide}\ and\ \citenamefont
  {Hemati}(2023)}]{heide2023optimization}%
  \BibitemOpen
  \bibfield  {author} {\bibinfo {author} {\bibfnamefont {A.~L.}\ \bibnamefont
  {Heide}}\ and\ \bibinfo {author} {\bibfnamefont {M.}~\bibnamefont {Hemati}},\
  }\bibfield  {title} {\bibinfo {title} {An optimization framework for
  analyzing nonlinear stability due to sparse finite-amplitude perturbations},\
  }in\ \href@noop {} {\emph {\bibinfo {booktitle} {AIAA AVIATION 2023 Forum}}}\
  (\bibinfo {year} {2023})\ p.\ \bibinfo {pages} {4258}\BibitemShut {NoStop}%
\bibitem [{\citenamefont {Heide}\ and\ \citenamefont
  {Hemati}(2024)}]{heideARXIV2024}%
  \BibitemOpen
  \bibfield  {author} {\bibinfo {author} {\bibfnamefont {A.~L.}\ \bibnamefont
  {Heide}}\ and\ \bibinfo {author} {\bibfnamefont {M.~S.}\ \bibnamefont
  {Hemati}},\ }\bibfield  {title} {\bibinfo {title} {An optimization framework
  for analyzing nonlinear stability due to sparse finite-amplitude
  perturbations},\ }\href@noop {} {\bibfield  {journal} {\bibinfo  {journal}
  {pre-print arXiv:2311.09507}\ } (\bibinfo {year} {2024})}\BibitemShut
  {NoStop}%
\bibitem [{Note1()}]{Note1}%
  \BibitemOpen
  \bibinfo {note} {Here we work with complex variables because we will work
  with operators formulated using spectral methods in later sections. For real
  variables, the methods will be analogous to the ones derived
  here.}\BibitemShut {Stop}%
\bibitem [{\citenamefont {Moghaddam}\ \emph {et~al.}(2006)\citenamefont
  {Moghaddam}, \citenamefont {Weiss},\ and\ \citenamefont
  {Avidan}}]{moghaddam}%
  \BibitemOpen
  \bibfield  {author} {\bibinfo {author} {\bibfnamefont {B.}~\bibnamefont
  {Moghaddam}}, \bibinfo {author} {\bibfnamefont {Y.}~\bibnamefont {Weiss}},\
  and\ \bibinfo {author} {\bibfnamefont {S.}~\bibnamefont {Avidan}},\
  }\bibfield  {title} {\bibinfo {title} {Generalized spectral bounds for sparse
  {LDA}},\ }\href@noop {} {\bibfield  {journal} {\bibinfo  {journal}
  {Proceedings of the 23rd International Conference on Machine Learning}\ }
  (\bibinfo {year} {2006})}\BibitemShut {NoStop}%
\bibitem [{\citenamefont {Kuleshov}(2013{\natexlab{a}})}]{kuleshov2013}%
  \BibitemOpen
  \bibfield  {author} {\bibinfo {author} {\bibfnamefont {V.}~\bibnamefont
  {Kuleshov}},\ }\bibfield  {title} {\bibinfo {title} {Fast algorithms for
  sparse principal component analysis based on {R}ayleigh quotient iteration},\
  }\href@noop {} {\bibfield  {journal} {\bibinfo  {journal} {Proceedings of the
  30th International Conference on Machine Learning, PMLR}\ }\textbf {\bibinfo
  {volume} {28}},\ \bibinfo {pages} {1418} (\bibinfo {year}
  {2013}{\natexlab{a}})}\BibitemShut {NoStop}%
\bibitem [{\citenamefont {Skene}\ \emph
  {et~al.}(2022{\natexlab{b}})\citenamefont {Skene}, \citenamefont {Yeh},
  \citenamefont {Schmid},\ and\ \citenamefont {Taira}}]{skene}%
  \BibitemOpen
  \bibfield  {author} {\bibinfo {author} {\bibfnamefont {C.~S.}\ \bibnamefont
  {Skene}}, \bibinfo {author} {\bibfnamefont {C.-A.}\ \bibnamefont {Yeh}},
  \bibinfo {author} {\bibfnamefont {P.~J.}\ \bibnamefont {Schmid}},\ and\
  \bibinfo {author} {\bibfnamefont {K.}~\bibnamefont {Taira}},\ }\bibfield
  {title} {\bibinfo {title} {Sparsifying the resolvent forcing mode via
  gradient-based optimization},\ }\href@noop {} {\bibfield  {journal} {\bibinfo
   {journal} {Journal of Fluid Mechanics}\ }\textbf {\bibinfo {volume} {944}}
  (\bibinfo {year} {2022}{\natexlab{b}})}\BibitemShut {NoStop}%
\bibitem [{\citenamefont {Lopez-Doriga}\ \emph {et~al.}(2022)\citenamefont
  {Lopez-Doriga}, \citenamefont {Ballouz}, \citenamefont {Bae},\ and\
  \citenamefont {Dawson}}]{lopez}%
  \BibitemOpen
  \bibfield  {author} {\bibinfo {author} {\bibfnamefont {B.}~\bibnamefont
  {Lopez-Doriga}}, \bibinfo {author} {\bibfnamefont {E.}~\bibnamefont
  {Ballouz}}, \bibinfo {author} {\bibfnamefont {H.~J.}\ \bibnamefont {Bae}},\
  and\ \bibinfo {author} {\bibfnamefont {S.~T.~M.}\ \bibnamefont {Dawson}},\
  }\bibfield  {title} {\bibinfo {title} {A sparsity-promoting resolvent
  analysis for the identification of spatiotemporally-localized amplification
  mechanisms},\ }\href@noop {} {\bibfield  {journal} {\bibinfo  {journal}
  {arxiv:2212.02741}\ } (\bibinfo {year} {2022})}\BibitemShut {NoStop}%
\bibitem [{\citenamefont {Kuleshov}(2013{\natexlab{b}})}]{kuleshov}%
  \BibitemOpen
  \bibfield  {author} {\bibinfo {author} {\bibfnamefont {V.}~\bibnamefont
  {Kuleshov}},\ }\bibfield  {title} {\bibinfo {title} {Fast algorithms for
  sparse principal component analysis based on {R}ayleigh quotient iteration},\
  }\href@noop {} {\bibfield  {journal} {\bibinfo  {journal} {Proceedings of the
  30th International Conference on Machine Learning}\ } (\bibinfo {year}
  {2013}{\natexlab{b}})}\BibitemShut {NoStop}%
\bibitem [{\citenamefont {Parlett}(1998)}]{parlett}%
  \BibitemOpen
  \bibfield  {author} {\bibinfo {author} {\bibfnamefont {B.~N.}\ \bibnamefont
  {Parlett}},\ }\href@noop {} {\emph {\bibinfo {title} {The Symmetric
  Eigenvalue Problem}}}\ (\bibinfo  {publisher} {Society for Industrial and
  Applied Mathematics},\ \bibinfo {address} {Philadelphia},\ \bibinfo {year}
  {1998})\BibitemShut {NoStop}%
\bibitem [{\citenamefont {Journ\'ee}\ \emph {et~al.}(2010)\citenamefont
  {Journ\'ee}, \citenamefont {Nesterov}, \citenamefont {Richt\'arik},\ and\
  \citenamefont {Sepulchre}}]{journee2010}%
  \BibitemOpen
  \bibfield  {author} {\bibinfo {author} {\bibfnamefont {M.}~\bibnamefont
  {Journ\'ee}}, \bibinfo {author} {\bibfnamefont {Y.}~\bibnamefont {Nesterov}},
  \bibinfo {author} {\bibfnamefont {P.}~\bibnamefont {Richt\'arik}},\ and\
  \bibinfo {author} {\bibfnamefont {R.}~\bibnamefont {Sepulchre}},\ }\bibfield
  {title} {\bibinfo {title} {Generalized power method for sparse principal
  component analysis},\ }\href@noop {} {\bibfield  {journal} {\bibinfo
  {journal} {Journal of Machine Learning Research}\ }\textbf {\bibinfo {volume}
  {11}},\ \bibinfo {pages} {517–553} (\bibinfo {year} {2010})}\BibitemShut
  {NoStop}%
\bibitem [{\citenamefont {Foures}\ \emph {et~al.}(2014)\citenamefont {Foures},
  \citenamefont {Caulfield},\ and\ \citenamefont {Schmid}}]{foures2014optimal}%
  \BibitemOpen
  \bibfield  {author} {\bibinfo {author} {\bibfnamefont {D.~P.}\ \bibnamefont
  {Foures}}, \bibinfo {author} {\bibfnamefont {C.~P.}\ \bibnamefont
  {Caulfield}},\ and\ \bibinfo {author} {\bibfnamefont {P.~J.}\ \bibnamefont
  {Schmid}},\ }\bibfield  {title} {\bibinfo {title} {Optimal mixing in
  two-dimensional plane poiseuille flow at finite p{\'e}clet number},\
  }\href@noop {} {\bibfield  {journal} {\bibinfo  {journal} {Journal of Fluid
  Mechanics}\ }\textbf {\bibinfo {volume} {748}},\ \bibinfo {pages} {241}
  (\bibinfo {year} {2014})}\BibitemShut {NoStop}%
\end{thebibliography}%

% \clearpage
% \input{resolvent.tex}

\end{document}